\documentclass[aps,prl, twocolumn, superscriptaddress]{revtex4-2}
\usepackage{amsmath,amsfonts,amssymb}
\usepackage{hyperref}
\usepackage{graphicx}
\usepackage{blindtext}
\usepackage{isotope}
\usepackage{dcolumn}
\usepackage{bm}
\usepackage[mathlines]{lineno}
\usepackage[utf8]{inputenc} 
\usepackage[T1]{fontenc}
\usepackage{threeparttable} 
\usepackage{mathptmx}
\usepackage{etoolbox}
\usepackage[usenames,dvipsnames]{color}

\usepackage{epstopdf}
\usepackage{filecontents}
\usepackage{scalerel}
\usepackage{dcolumn}
\usepackage{braket}
\usepackage{epstopdf}
\usepackage{filecontents}
\usepackage{scalerel}
\usepackage{dcolumn}
\usepackage{booktabs}       
\usepackage{threeparttable} 
\usepackage{graphicx}       
\usepackage{siunitx}
\usepackage{caption}
\usepackage{dcolumn}
\usepackage{bm}
\usepackage{braket}
\usepackage[mathlines]{lineno}
\usepackage[utf8]{inputenc} 
\usepackage[T1]{fontenc}
\usepackage{mathptmx}
\usepackage{etoolbox}
\usepackage[usenames,dvipsnames]{color}
\newcommand{\dd}{\mathrm{d}}

\newcommand{\ee}{\mathrm{e}}

\begin{document}

\preprint{PRL/123-QED}
\title{Enhancement of Indistinguishable Photon Emission from a GaAs Quantum Dot via Charge Noise Suppression}

\author{Priyabrata Mudi,\textsuperscript{\textdagger}}
\email{mudi@tu-berlin.de}
\author{Avijit Barua}
\thanks{These authors contributed equally to this work.}
\author{Kartik Gaur}%
\affiliation{Institut für Physik und Astronomie,
Technische Universität Berlin, Hardenbergstraße 36, D-10623 Berlin, Germany
}
\author{Steffen Wilksen}
\author{Alexander Steinhoff}
\affiliation{Institute for Physics, Carl von Ossietzky Universität Oldenburg, 26129 Oldenburg, Germany}%
\author{Setthanat Wijitpatima}%
\author{Sarthak Tripathi}%
\affiliation{Institut für Physik und Astronomie,
Technische Universität Berlin, Hardenbergstraße 36, D-10623 Berlin, Germany
}%
\author{Julian Ritzmann}%
\author{Andreas D. Wieck}%
\affiliation{
Lehrstuhl für Angewandte Festkörperphysik, Ruhr-Universität Bochum, DE-44780 Bochum, Germany
}%
\author{Sven Rodt}%
\affiliation{Institut für Physik und Astronomie,
Technische Universität Berlin, Hardenbergstraße 36, D-10623 Berlin, Germany
}%
\author{Christopher Gies}
\affiliation{Institute for Physics, Carl von Ossietzky Universität Oldenburg, 26129 Oldenburg, Germany}%
\author{Arne Ludwig}
\affiliation{
Lehrstuhl für Angewandte Festkörperphysik, Ruhr-Universität Bochum, DE-44780 Bochum, Germany
}%
\author{Stephan Reitzenstein}
\email{stephan.reitzenstein@physik.tu-berlin.de}
\affiliation{Institut für Physik und Astronomie,
Technische Universität Berlin, Hardenbergstraße 36, D-10623 Berlin, Germany
}%

\date{\today}

\begin{abstract}
The generation of indistinguishable single photons is a fundamental requirement for future quantum technologies, particularly in quantum repeater networks and for distributed quantum computing based on entanglement distribution. However, spectral jitter, often induced by charge noise in epitaxial quantum dots, leads to exciton dephasing, thereby limiting their practical usage in quantum applications. We present a straightforward approach to mitigate charge noise-induced decoherence in droplet-etched GaAs quantum dots embedded in an n-i-p diode structure and integrated deterministically into an electrically contacted circular Bragg grating resonator for emission enhancement. The quantum device allows for the stabilization of the charge environment by applying an external electrical field while producing a photon extraction efficiency of $ (37 \pm 2) \%$. Hong-Ou-Mandel two-photon interference measurements reveal a strong voltage dependence of the exciton dephasing time and interference visibility on the applied bias in excellent agreement with our theoretical predictions. Notably, the reduction in visibility from a maximum, charge stabilized corrected value of 97\% at the optimum bias point follows an inverse square dependence ($\propto 1/I^2$) with increasing diode current ($I$) in forward direction. Under a quasi-resonant excitation scheme, we achieve a maximum exciton dephasing time $(T_2^*)$ of approximately $(6.8\pm0.5)$ ns, reaching nearly the Fourier limit ($T_2 = 2T_1$) without the need for complex echo schemes like Ramsey or Carr–Purcell–Meiboom–Gill sequences. These findings are consistent with theoretical predictions from rate equation modeling and quantum optical analysis as well as voltage-dependent linewidth measurements, demonstrating optimized electrical control of exciton dephasing.
\end{abstract}

\maketitle
The foundation principle of quantum communication and distributed quantum computing relies on the generation of indistinguishable single photons from quantum light sources \cite{lu2014push, brito2020statistical,avron2021quantum}. Epitaxially grown quantum dots (QDs), particularly self-assembled In(Ga)As QDs, have shown excellent performance as on-demand single-photon emitters \cite{heindel2023quantum, hornung2024highly, gschrey2015highly, Maass_Barua2025, kuhlmann2015transform}, but their use is limited by large fine-structure splitting (FSS) \cite{seguin2005size} and spin decoherence arising from nuclear spin noise due to high spin \isotope[115]{In} and \isotope[113]{In} isotopes \cite{urbaszek2013nuclear, gangloff2019quantum}. In this context, GaAs QDs grown via droplet etching in an AlGaAs matrix offer a promising alternative \cite{heyn2009highly}.  Owing to their high structural symmetry, GaAs QDs exhibit significantly reduced FSS \cite{huo2013ultra} and their emission in the 700-800 nm range overlaps with the D1/D2 transitions of rubidium atoms. This spectral alignment enables potential integration with atomic memory platforms \cite{jahn2015artificial, huang2017electrically, keil2017solid}. Additionally, the absence of high-spin indium isotopes improves spin coherence \cite{nguyen2023enhanced}. These properties have enabled GaAs QDs to emit single photons with enhanced spin coherence \cite{nguyen2023enhanced}, improved indistinguishability \cite{zhai2022quantum}, and high entanglement fidelity \cite{huber2018strain}. Despite this, GaAs QDs face challenges including spectral diffusion and blinking due to charge noise \cite{ha2015size,beguin2018demand,jahn2015artificial}. Although embedding QDs in an n-i-p diode can stabilize the charge environment, doping AlGaAs layers often introduces DX centers \cite{mooney1990deep,munoz1993techniques}, which contribute to additional charge noise. Nevertheless, recent advancements have shown that reducing Al composition in epitaxial gate suppresses DX center formation and enables blinking-free operation in the majority of QDs \cite{zhai2020low}.

\par In this work, we investigate two exemplary droplet-etched GaAs QDs, QD1 and QD2, embedded in an n-i-p diode structure within an AlGaAs matrix \cite{zhai2020low, babin2021charge, sm} and deterministically integrated into an electrically contacted circular Bragg grating (eCBG) cavity \cite{wijitpatima2024bright}. These nanophotonic structures, incorporating a nano-ridge for electrical connectivity, provide high extraction efficiency up to $\left(37\pm2\right)\%$ \cite{sm} and broad voltage tunability. Before integration, approximately 70\% of such QDs exhibit stable, blinking-free emission on a millisecond timescale as confirmed by the second-order autocorrelation measurements under continuous-wave (cw) excitation \cite{zhai2020low}, with reported two-photon interference visibilities reaching 93\% without temporal or spectral filtering \cite{zhai2022quantum}. To study the influence of diode current-introduced charge noise, QDs susceptible to blinking are preselected via cathodoluminescence (CL) mapping at 20 K and integrated using marker-based deterministic electron beam lithography \cite{li2023scalable,sm}. Focusing on a noisy QD located near the wafer edge, we observe a pronounced voltage dependence of the Hong-Ou-Mandel (HOM) two-photon interference visibility, coherence time ($T_2$), and dephasing time ($T_2^*$). At optimal bias, we achieve $T_2 \approx 2T_1$ and $T_2^* = (6.8 \pm 0.5) $ ns under quasi-resonant excitation. The observed degradation in visibility at higher bias voltages is accurately reproduced by a quantum optical model incorporating diode current ($I$) induced charge noise, consistent with a $1/I^2$ dependence.

\begin{figure}[!h]
\includegraphics[width=0.45\textwidth]{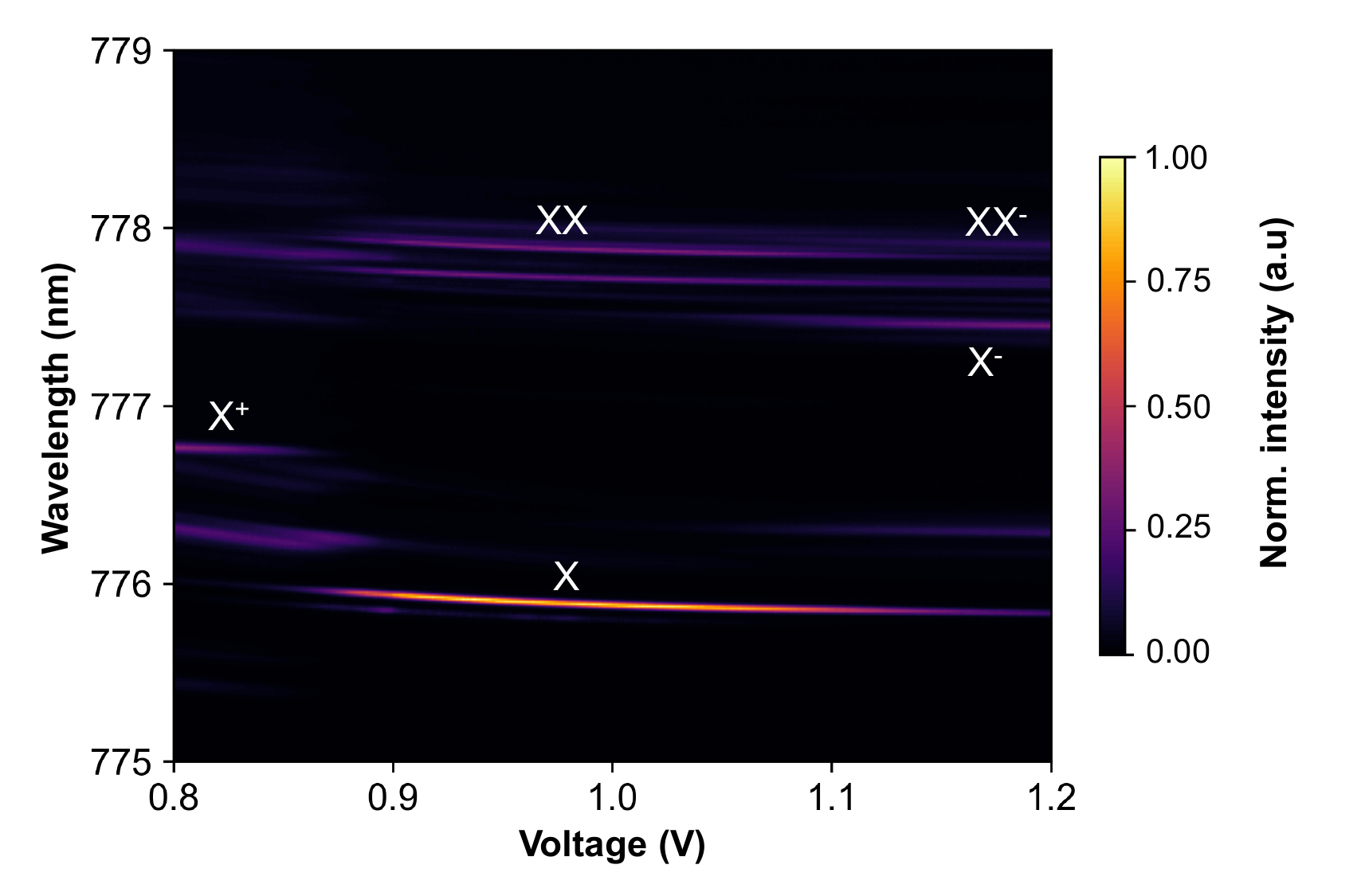}
\centering
\caption{$\mu$-PL spectra of QD1 as a function of gate voltage under quasi-resonant excitation, showing the emergence and suppression of distinct emission lines.}\label{Fig 1}
\vspace{-10pt}
\end{figure}

\par Following deterministic nano-fabrication, the QD sample is characterized by recording micro-photoluminescence ($\mu$-PL) spectra as a function of gate voltage ($V_G$), using a longitudinal optical (LO) phonon-assisted quasi-resonant excitation scheme, in which the laser energy is blue-detuned by one LO phonon energy (36 meV) from the QD emission. The measured spectra, as illustrated in Fig.~\ref{Fig 1}, exhibit distinct QD emission lines corresponding to various excitonic complexes and charge states, which appear and disappear as the bias voltage is varied. This phenomenon is attributed to changes in occupation probability of different charge state configurations \cite{warburton2000optical}.  The excitonic-biexcitonic ($X$-$XX$) features appear within the voltage range of 0.8–1.2~V, and the nature of the emission lines is identified through power- and polarization-resolved measurements. A typical fine-structure splitting (FSS) of 1–2 GHz is observed for both QD1 and QD2, consistent with values reported for other QDs in this sample \cite{zhai2020low} and in the literature for (001)-oriented wafers \cite{huo2013ultra, liu2019solid}. Notably, the electrical tuning of the present sample allows a wide control of the charge configuration ranging from $X^{2+}$ to $X^{8-}$; however, the present work focuses on neutral excitons ($X^0$), which are seen to be more susceptible to charge noise \cite{undeutsch2025electric} and are particularly relevant for the generation of polarization-entangled photon pairs. 

\par
\begin{figure}
\includegraphics[width=0.49\textwidth]{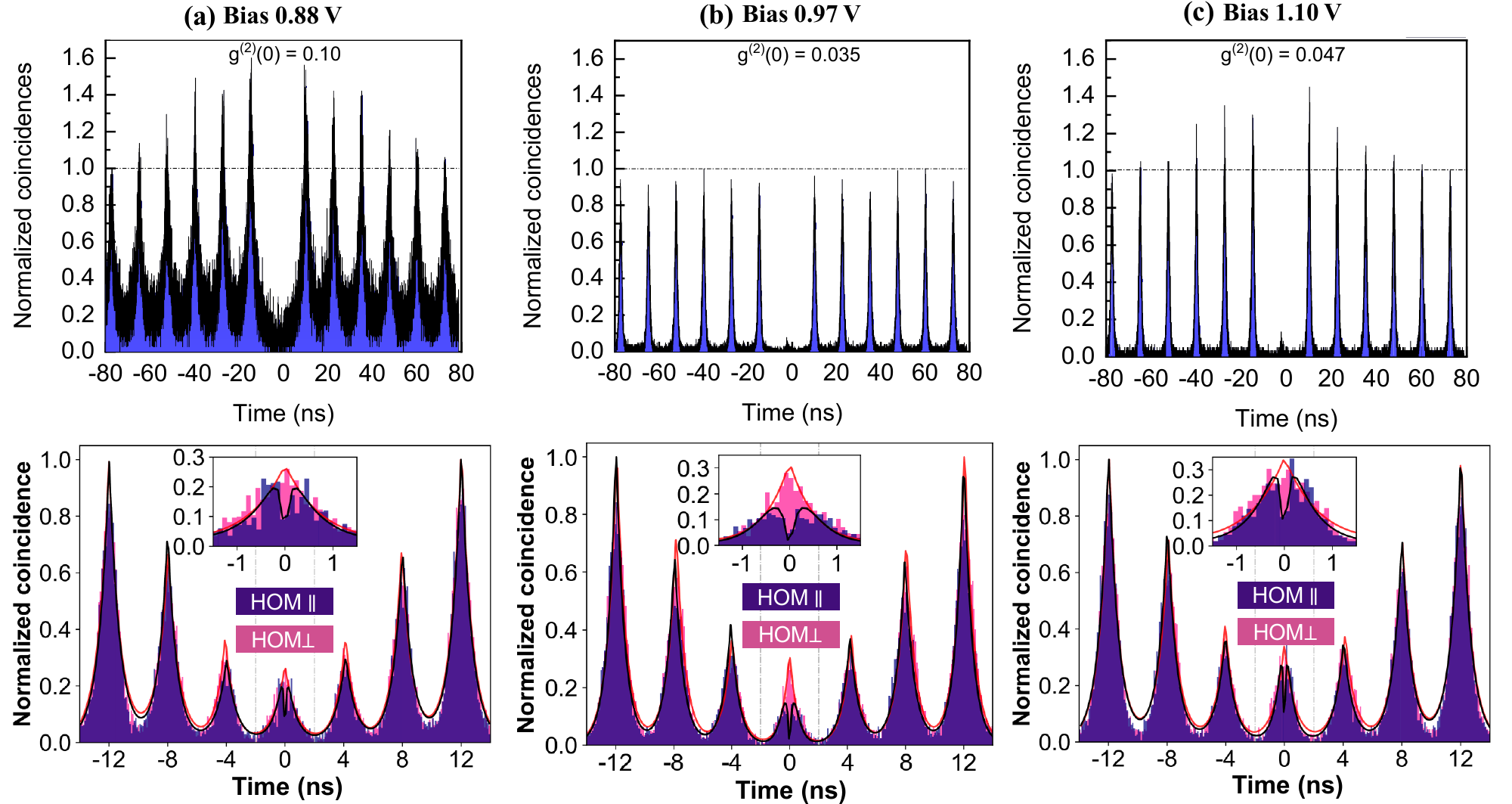}
\centering
\caption{Second-order autocorrelation measurements (top panel) and photon coincidence histograms from Hong-Ou-Mandel interference (bottom panel) of the QD1 $X^0$ line under quasi-resonant excitation at different bias voltages: (a) 0.88 V, (b) 0.97 V, and (c) 1.10 V }\label{Fig 2} 
\vspace{-10 pt}
\end{figure}

 In order to investigate the effect of charge noise, second-order autocorrelation measurements are performed at various bias voltages using a pulsed (80 MHz) Ti:Sapphire laser in ps-mode under quasi-resonant excitation at 759 nm (LO phonon assisted excitation), as illustrated in Fig.~\ref{Fig 2} [top panel]. Notably, despite extensive efforts, no QD emission was detected under strictly resonant excitation in our devices. This is likely due to the polarization selectivity of the eCBG cavity; however, a detailed investigation lies beyond the scope of the present study. As a result, all measurements in this study are performed under slightly detuned quasi-resonant excitation, despite its inherent limitations. The coincidence histograms are recorded using a Hanbury-Brown and Twiss (HBT) setup to inspect the photon statistics and to identify potential blinking. At a bias voltage of 0.88 V, the coincidence peaks at shorter delays exhibit an increased number of coincidences, a potential signature of inherent charge fluctuation or QD blinking \cite{hopfmann2021heralded}. The behavior is due to fluctuation in the local charge environment, leading to transient trapping and detrapping of electrons and holes that perturb the QD emission \cite{hopfmann2021heralded}. At a slightly higher bias (0.97 V), however, the height of the coincidence peaks at both short and long delays becomes nearly identical, indicating effective suppression of charge noise and stabilization of QD emission. The absence of bunching confirms an optimal voltage range of 0.925 V to 1.020 V where the charge environment remains stable, eliminating blinking effects. To further verify these results, similar measurements are performed under weak cw excitation (600 nW) that did not show blinking up to 400 ns (see Fig.~S1). Nevertheless, at 1.05~V, bunching in the envelope function re-emerges, which is attributed to charge noise introduced by external currents.

 \begin{figure*}
\includegraphics[width=0.85\textwidth]{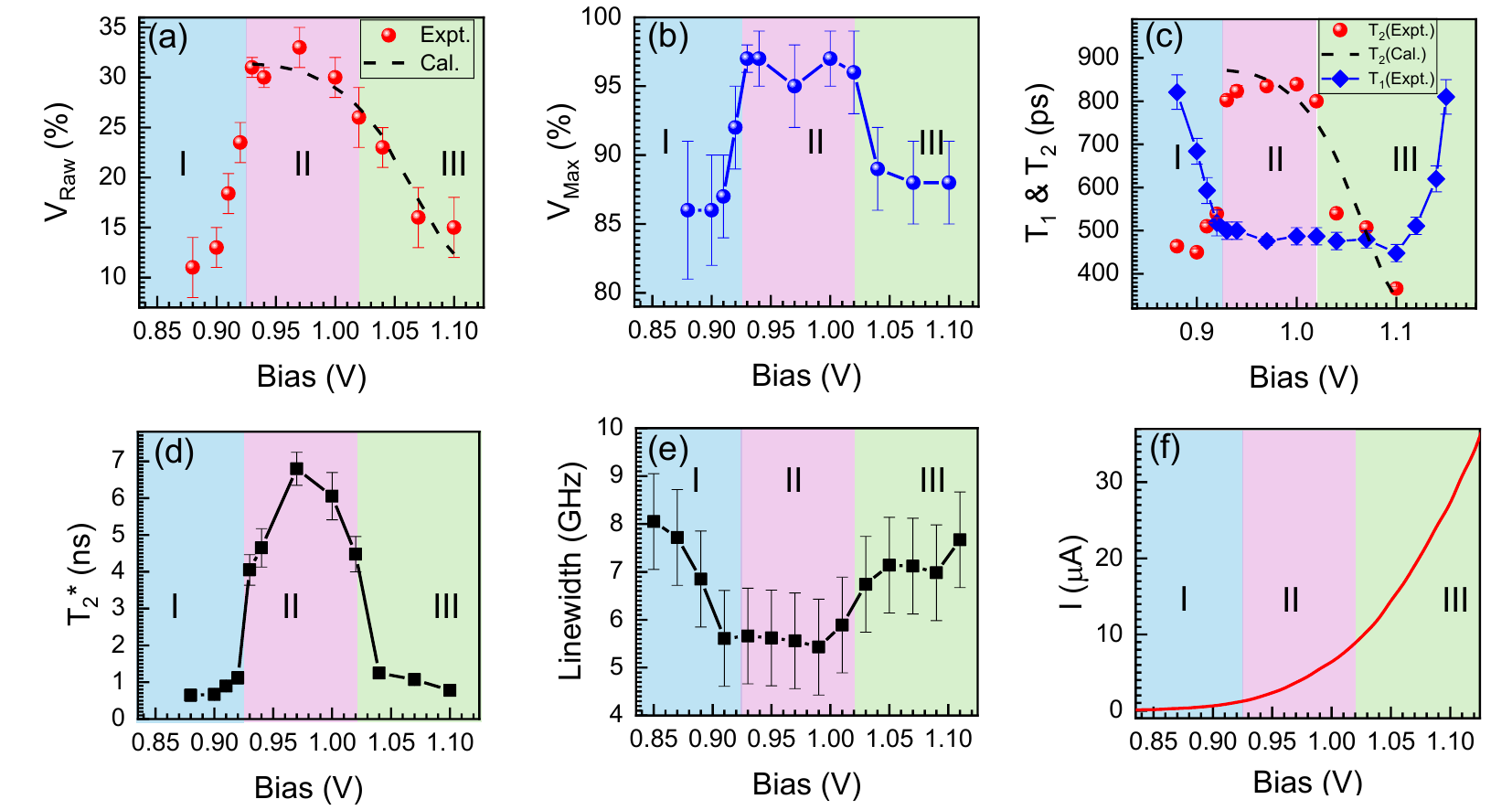}
\centering
\caption{Bias-dependent properties of the QD1 $X^0$ line under quasi-resonant excitation: (a) Raw two-photon visibility experimentally measured (symbol) and theoretically modeled (dashed line), (b) time-filtered maximum two-photon visibility, (c)  exciton lifetime ($T_1$) and coherence time ($T_2$)- both experimental and theoretical (dashed line), (d) pure dephasing time ($T_2^*$), and (e) emission linewidth (inhomogeneous) all plotted as a function of applied bias voltage.  (f) Current-voltage (I-V) characteristics of the n-i-p diode.} \label{Fig 4} 
\vspace{-10pt}
\end{figure*}

\par To understand the underlying mechanisms responsible for the appearance of photon bunching at both lower and higher bias voltages, as well as the optimal intermediate bias voltage range where charge noise and bunching are effectively suppressed, HOM two-photon interference measurements are performed at varying bias voltages. These measurements reveal a strong bias dependence, as demonstrated in Fig.~\ref{Fig 2} [bottom panel]. Notably, at the onset of the bias plateau for the QD1 $X^0$ line (0.85 V), the exciton lifetime is significantly prolonged (from 0.5 to 2.1 ns) due to band bending and a reduced overlap between the electron and hole wavefunctions \cite{undeutsch2025electric}. This effect results in overlapped peaks in the HOM coincidence histogram. Consequently, HOM interference measurements are only conducted within the (0.88–1.10) V bias range, where the exciton lifetime remains much lower than 4 ns (delay between two consecutive photons in the HOM experiments). Recently, Undeutsch et al. \cite{undeutsch2025electric} reported a non-monotonic variation of exciton lifetime affecting two-photon visibility in XX–X cascade decays. However, their study was restricted to bias voltages below 1 V due to luminescence quenching caused by single-electron charging. In contrast, our integration of QDs into eCBGs ensures bright $X^0$ emission along with bias control, allowing measurements beyond the onset of significant forward current at 1 V. The HOM two-photon visibility measurements show a distinctive volcano-shaped reduction in the central peak for co-polarized detection in both arms of the HOM interferometer. This volcano shape originates from the imperfect excitation scheme \cite{Michler2009}, which introduces unintended time jitter between consecutive photon arrivals at the HOM interferometer beam splitter. It can be observed that at 0.88 V [Fig.~\ref{Fig 2}(a) (bottom) inset] the central peak in the photon coincidence histogram exhibits nonzero events even at zero delay because of limited two-photon visibility. However, at 0.97 V [Fig.~\ref{Fig 2}(b) (bottom) inset], the number of events at zero delay is significantly reduced, signifying improved visibility, which aligns with the optimal bias range identified in the HBT measurements. At a higher voltage of 1.10 V [Fig.~\ref{Fig 2}(c) (bottom) inset], the number of zero-delay correlation events increases again, indicating a deterioration in two-photon visibility. Such a bias dependence of visibility is not observed for $XX$ \cite{undeutsch2025electric} or $X^{-}$ lines \cite{wijitpatima2024bright}.

\par Fig.~\ref{Fig 4}(a) shows the raw two-photon visibility of the QD1 $X^0$ transition in the bias voltage range (0.88–1.10) V. The bias dependence of the maximum visibility ($V$) and coherence time ($T_2$) presented in Fig.~\ref{Fig 4}(b) \& (c) are obtained from the co-polarized HOM coincidence histogram by fitting the central peak ($C_{HOM, \parallel}(\tau)$) \cite{kimtwo2025, bylander2003interference, sm} at each bias voltage. The results obtained indicate that the QD system can be categorized into three distinct regions depending on the applied bias: (i) region I (0.850 V – 0.925 V), (ii) region II (0.925 V – 1.020 V), and (iii) region III (1.02 V – 1.10 V).  Region I is located near the edge of the charge plateau, where charge fluctuations in the local environment of the QD induce shifts in the Fermi level near the emitter, leading to the stochastic capture and release of a hole from the Fermi sea. Consequently, the QD periodically switches between the $X^0$ and $X^+$ states, manifesting as blinking in both photoluminescence and autocorrelation measurements. This charge-induced instability is reflected in the linewidth measurements shown in Fig.~\ref{Fig 4}(e), where the linewidth is largest at the lowest voltage and systematically decreases as the bias voltage increases. A similar linewidth broadening trend has been previously reported in Ref. \cite{zhai2020low}, corroborating the presence of charge noise at low bias voltages. Moreover, both the two-photon visibility and coherence time exhibit reduced values ($\approx 12 \%$ \& $ 450$ ps respectively) in this region, consistent with the impact of blinking on the photon indistinguishability \cite{jons2017two}. Notably, the longer lifetime in this region also contributes to the degradation of two-photon visibility, as decoherence mechanisms are more likely to act before photon emission occurs. To further investigate the blinking phenomenon, we expanded a rate equation model introduced in Ref. \cite{sallen2010subnanosecond, sm} that predicts the behavior of the second-order autocorrelation function under cw excitation. The model yields the blinking on-off ratio parameter ($\beta$), which follows a similar trend as two-photon visibility \cite{sm}, confirming that the charge environment remains unstable in region I, where pronounced blinking persists.

\par  As the bias voltage approaches region II, the charge environment becomes more stabilized, effectively minimizing the charge noise and spectral fluctuations. This stabilization is critical for the QD's quantum optical properties, as it suppresses charge noise-induced decoherence mechanisms.  Notably, diode current in this region is considerably small and its change does not initiate any additional charge noise. As a consequence of the improved charge stability and suppressed blinking \cite{sm}, both the two-photon visibility and coherence time reach their optimal values of approximately $\left( 33 \pm 3\right)\%$ \& $ \left( 835 \pm 12 \right)$ ps, respectively. The maximum two-photon visibility at zero time achieves an impressive value of $97 \%$ at 0.97 V under quasi-resonant excitation, which is comparable to the value of best-performing low noise GaAs QDs free from blinking effect under strictly resonant excitation\cite{zhai2022quantum}. This enhanced visibility suggests that the charge noise in a non-ideal GaAs QD, which typically deteriorates photon indistinguishability, is effectively suppressed within the voltage range of region II. At the same bias voltage, coherence-to-lifetime ratio, $T_2/T_1$ reaches values as high as $1.8 \pm 0.1 $, which is remarkably close to the Fourier limit of $T_2 = 2T_1$.  The linewidth in region II is almost independent of the bias voltage. Notably, the linewidth presented here is broader than its intrinsic value due to the limitations of the quasi-resonant excitation scheme. Under strictly resonant excitation, similar QDs have been observed to approach the transform-limited linewidth \cite{zhai2020low}. The high coherence time measured in this work further supports this interpretation, indicating that the system is near its fundamental quantum coherence limit under electrical charge control. This also signifies a strong reduction in pure dephasing as the estimated pure dephasing time reaches as high as  $(6.8\pm 0.5)$ ns without the need for any complicated pulse sequence techniques \cite{nguyen2023enhanced}, demonstrating the potential for achieving long-lived quantum coherence under relatively simple experimental conditions beneficial for real-world applications. 

\par At higher bias voltages, corresponding to region III, the QD emission intensity remains relatively stable, indicating that it is still well within the charge plateau and far away from $X^0$ to $X^-$ transition edge. However, a significant increase in the diode current is observed in this region, introducing additional charge noise from externally injected electrons and holes in the active region leading to a higher blinking \cite{sm}. Dobrzanski et al.\cite{dobrzanski2004low} have studied the relationship between diode current (I) and low-frequency $1/f$-type charge noise (S) in QDs embedded within n-i-p structure, demonstrating that for current values $\leq 10^{-4} A$, as in our case, charge noise follows the relation $S \propto I^2$ \cite{sm}. To model the current dependence of the indistinguishability, we first calculated the two-time correlation function $C(t,\tau) = \braket{\hat{\sigma}^+(t+\tau) \hat{\sigma}^-(t)}$, where $\hat{\sigma}^+$ ($\hat{\sigma}^-$) is the raising (lowering) operator of the QD transition. In addition to the two levels involved in the optical transition ($\ket{\mathrm{g}}$ and $\ket{\mathrm{e}}$), a pump state $\ket{\mathrm{p}}$ is included to account for timing jitter introduced by quasi-resonant excitation. The correlation function is obtained by using the adjoint master equation formalism \cite{breuer2002theory} and the quantum regression theorem \cite{carmichael2013statistical}, yielding \cite{sm},
\begin{equation}
    C(t, \tau) =
    \begin{cases}
        \frac{\gamma_\mathrm{p}}{\gamma_\mathrm{rad} - \gamma_\mathrm{p}} 
                \left( \ee^{-\gamma_\mathrm{p}t} - \ee^{-\gamma_\mathrm{rad}t} \right)
                \ee^{-\frac{1}{2} (\gamma_\mathrm{rad} + \gamma_\mathrm{deph})\tau} 
                 &, \gamma_\mathrm{p} \neq \gamma_\mathrm{rad} \\
        \gamma_\mathrm{rad} t \ee^{-\gamma_\mathrm{rad}t} \ee^{-\frac{1}{2} (\gamma_\mathrm{rad} + \gamma_\mathrm{deph})\tau} & ,\gamma_\mathrm{p} = \gamma_\mathrm{rad}.
    \end{cases}
\end{equation}
Here, $\gamma_\mathrm{p}$ denotes the relaxation rate from the pump state to the radiative state, $\gamma_\mathrm{rad}$ is the radiative decay rate. The pure dephasing rate is given by $\gamma_\mathrm{deph} = \alpha I^2 + \gamma_\mathrm{pure}^{(0)}$
with the proportionality constant $\alpha$ and a small bias-independent offset $\gamma_\mathrm{pure}^{(0)}$. With these parameters, the two-photon indistinguishability can be calculated from the correlation function via \cite{kaer2013microscopic,steinhoff2025impact}
\begin{equation}\label{eq:v_integrals}
        V_\mathrm{raw} = \frac{\int_0^\infty \dd t \int_0^\infty \dd \tau \left|C(t,\tau)\right|^2}{\int_0^\infty \dd t \int_0^\infty \dd \tau \, C(t+\tau, 0) C(t,0)}.
\end{equation}
Using $\alpha$, $\gamma_\mathrm{p}$ and $\gamma_\mathrm{pure}^{(0)}$ as fitting parameters, the model accurately reproduces the measured values of two photon visibility in regions II and III (see Fig.~\ref{Fig 4}\,(a)). Applying the same parameters, the measured coherence time $T_2$ also is well described by $T_2 = 2 / (\alpha I^2 + \gamma_\mathrm{deph}^{(0)} + \gamma_\mathrm{rad})$ (see Fig.~\ref{Fig 4}\,(c)). Notably, this quantum optical model is only valid in regions II and III where diode current exhibits a clear voltage dependence. In contrast, region I is dominated by charge state blinking, which introduces linewidth broadening through a distinct mechanism. Extending the model to region I would require microscopic modeling to obtain bias-dependent blinking rates of this specific QD, which lies beyond the scope of this paper. Overall, the model captures both the observed degradation in two-photon visibility and coherence time with increasing voltage in region III. Additionally, the linewidth exhibits a systematic increase with bias in region III, further confirming the presence of charge noise. Note that the mechanism of linewidth broadening in region III is different from that in region I, as the bias is not yet high enough to reach the other end of the charge plateau. The aforementioned effect is also responsible for the reappearance of bunching in the envelope function of HBT measurements. Additionally, the linewidth exhibits a systematic increase with bias in region III, further confirming the presence of charge noise. 

\par In summary, the present work furnishes an approach to suppress charge noise-induced decoherence in droplet-etched GaAs QDs embedded within an n-i-p diode structure and integrated into an eCBG nanocavity for electrical control and emission enhancement. By applying an external electric field, the charge environment is stabilized while maintaining a photon extraction efficiency of $(37\pm2)\%$. HBT and HOM two-photon interference measurements reveal a strong bias dependence of the single-photon purity, exciton dephasing time, and interference visibility. At an optimal bias voltage, the coherence time reaches 1.8 times the lifetime, approaching the Fourier limit ($T_2 = 2T_1$), and the pure dephasing time $T_2^*$ extends up to $(6.8\pm0.5)$ ns, all under simple quasi-resonant excitation. These findings demonstrate that by maximizing the trade-off between charge stability and diode current, they provide a workable method to reduce charge noise in GaAs QDs, highlighting its promise for scalable photonic quantum technology.  

\begin{acknowledgments}
This work was financially supported by the German Federal Ministry of Education and Research (BMBF) through the projects 16KISQ012, 16KISQ014, and 16KIS2203, and by the Germany Research Foundation (DFG) through the project PhotonicQRC. A.L. acknowledges support by the BMFTR Project QR.N 16KIS2200, QuantERA BMFTR EQSOTIC 16KIS2061, and DFG ML4Q EXC 2004/1. The authors thank Prof. Richard J. Warburton for valuable scientific discussions and insightful comments. Additionally, the authors acknowledge VI Systems GmbH for their technical assistance with wire bonding.
\end{acknowledgments}

\bibliographystyle{unsrt}
\bibliography{References}

\end{document}